# ASTROD-GW: OVERVIEW AND PROGRESS

WEI-TOU NI
*Center for Gravitation and Cosmology (CGC), Department of Physics,
National Tsing Hua University, Hsinchu, Taiwan, 300, ROC*
*weitou@gmail.com*

In this paper, we present an overview of ASTROD-GW (ASTROD [Astrodynamical Space Test of Relativity using Optical Devices] optimized for Gravitational Wave [GW] detection) mission concept and its studies. ASTROD-GW is an optimization of ASTROD which focuses on low frequency gravitational wave detection. The detection sensitivity is shifted by a factor of 260 (52) towards longer wavelengths compared with that of NGO/eLISA (LISA). The mission consists of three spacecraft, each of which orbits near one of the Sun-Earth Lagrange points (L3, L4 and L5), such that the array forms an almost equilateral triangle. The 3 spacecraft range interferometrically with one another with an arm length of about 260 million kilometers. The orbits have been optimized resulting in arm length changes of less than $\pm 0.00015$ AU or, fractionally, less than $\pm 10^{-4}$ in twenty years, and relative Doppler velocities of the three spacecraft of less than $\pm 3$ m/s. In this paper, we present an overview of the mission covering: the scientific aims, the sensitivity spectrum, the basic orbit configuration, the simulation and optimization of the spacecraft orbits, the deployment of ASTROD-GW formation, TDI (Time Delay Interferometry) and the payload. The science goals are the detection of GWs from (i) Supermassive Black Holes; (ii) Extreme-Mass-Ratio Black Hole Inspirals; (iii) Intermediate-Mass Black Holes; (iv) Galactic Compact Binaries and (v) Relic Gravitational Wave Background. For the purposes of primordial GW detection, a six spacecraft formation would be needed to enable the correlated detection of stochastic GWs. A brief discussion of the six spacecraft orbit optimization is also presented.

**Contents**

**1. Introduction**

Historically the orbit and gravity observations/experiments in the solar-system have been important resources for the development of fundamental physical laws as the precision and accuracy are improved. It is so for both the developments of Newtonian world system and Einstein's general relativity. With the eminent improvement for the orbits and gravity measurements pending, we are in a historical epoch for a great stride in the development of fundamental laws. The gravitational field in the solar system is determined by three factors: the dynamic distribution of matter in the solar system; the dynamic distribution of matter outside the solar system (galactic, cosmological, etc.) and gravitational waves propagating through the solar system. Different relativistic/cosmological theories of gravity make different predictions of the solar-system gravitational field. Hence, precise measurements of the solar-system gravitational field test these relativistic theories, in addition to enabling gravitational wave (GW) observations, determination of the matter distribution in the solar-system and determination of the observable (testable) influence of our galaxy and cosmos. To measure the solar-system gravitational field, we measure/monitor distance between different natural and/or artificial celestial bodies. In the solar system, the equation of motion of a celestial body or a spacecraft is given by the astrodynamical equation

$$\mathbf{a} = \mathbf{a}_N + \mathbf{a}_{1PN} + \mathbf{a}_{2PN} + \mathbf{a}_{Gal\text{-}Cosm} + \mathbf{a}_{GW} + \mathbf{a}_{non\text{-}grav}, \tag{1}$$

where $\mathbf{a}$ is the acceleration of the celestial body or spacecraft, $\mathbf{a}_N$ is the acceleration due to Newtonian gravity, $\mathbf{a}_{1PN}$ the acceleration due to first post-Newtonian effects, $\mathbf{a}_{2PN}$ the acceleration due to second post-Newtonian effects, $\mathbf{a}_{Gal\text{-}Cosm}$ the acceleration due to Galactic and cosmological gravity, $\mathbf{a}_{GW}$ the acceleration due to GW, and $\mathbf{a}_{nongrav}$ the acceleration from all non-gravitational origins.[1] Distances between spacecraft depend critically on the solar-system gravity (including gravity induced by solar oscillations), underlying gravitational theory and incoming GWs. A precise measurement of these distances as a function of time will enable the cause of variation to be determined.

Certain orbit configurations are good for testing relativistic gravity; certain configurations are good for measuring solar parameters; certain are good for detecting gravitational waves. These factors are integral part of mission designs for various purposes.[1,2]

The basic concept of ASTROD (Astrodynamical Space Test of Relativity using Optical Devices) is to have a constellation of drag-free spacecraft navigate through the solar system and range with one another using optical devices to map the solar-system gravitational field, to measure related solar-system parameters, to test relativistic gravity, to observe solar g-mode oscillations, and to detect GWs.[1,2]

The distance determination of satellite laser ranging with two colors (two wavelengths) has reached millimeter accuracy. With the newer generation of lunar laser ranging,[3,4] the accuracy of lunar distance determination has also reached millimeter accuracy. On board timing accuracy of 3 ps (0.9 mm) has already achieved by the T2L2 (Time Transfer by Laser Link) event timer onboard Jason 2 satellite.[5,6] Based on these developments, the one-way ranging goal of ASTROD I (single spacecraft ranging in both ways with ground laser stations) -- the first mission of the ASTROD series -- is to millimeter accuracy. With this accuracy and extended ranges of 1 AU, the capability of probing the fundamental laws of spacetime and mapping the solar system gravity will be greatly enhanced.[7-11] Redshift measurement and comparison in the solar system together with Michelson-Morley and Kennedy-Thorndike experiments test the foundation of general relativity and metric theories of gravity, i.e., the Einstein Equivalence Principle (EEP).[12] OPTIS[13] is such a mission proposal in space. A possible combined ASTROD I and OPTIS mission is proposed to test both the dynamics and the EEP foundation of relativistic gravity in the solar system.[7,10,11]

In this paper, we address the issue of GW detection and give an overview of the ASTROD-GW (ASTROD [Astrodynamical Space Test of Relativity using Optical Devices] optimized for Gravitational Wave [GW] detection) mission concept and the progress in its study. ASTROD-GW is an optimization of ASTROD to focus on the goal of detection of GWs.[1,14-16] The scientific aim is focused on GW detection at low frequency. The mission orbits of the 3 spacecraft forming a nearly equilateral triangular array are chosen to be near the Sun-Earth Lagrange points L3, L4 and L5 (Fig. 1).[14-16] Due to its longer arm length compared to LISA[17] and NGO/eLISA,[18] ASTROD-GW has the best sensitivity goal in the lower part (100 nHz – 1mHz) of the low frequency band.[1,14-16]

Now we turn to the classification of GWs. Similar to frequency classification of electromagnetic waves to radio wave, millimeter wave, infrared, optical, ultraviolet, X-ray and γ-ray etc., we have the following complete Frequency Classification of Gravitational Waves (Fig. 2):[1,19-21]

(i) Ultra high frequency band (above 1 THz): Detection methods include Terahertz resonators, optical resonators, and ingenious methods yet to be invented.

(ii) Very high frequency band (100 kHz – 1 THz): Microwave resonator/wave guide detectors, optical interferometers and Gaussian beam detectors are sensitive to this band.

(iii) High frequency band (10 Hz – 100 kHz): Laser-interferometric ground detectors and low-temperature resonators are most sensitive to this band.

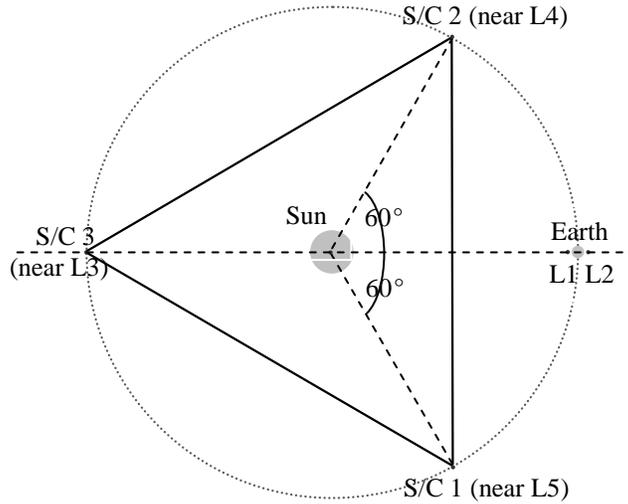

Fig. 1. Schematic of ASTROD-GW mission orbit design.

(iv) Middle frequency band (0.1 Hz – 10 Hz): Space interferometric detectors of short arm length (1000-100000 km) are most sensitive to this band.

(v) Low frequency band (100 nHz – 0.1 Hz): Laser-interferometer space detectors are most sensitive to this band.

(vi) Very low frequency band (300 pHz – 100 nHz): Pulsar timing observations are most sensitive to this band.

(vii) Ultra low frequency band (10 fHz – 300 pHz): Astrometry of quasar proper motions is most sensitive to this band.

(viii) Extremely low (Hubble) frequency band (1 aHz – 10 fHz): Cosmic microwave background experiments are most sensitive to this band.

(ix) Infra Hubble frequency band (below 1 aHz): Inflationary cosmological models give strengths of GWs in this band. They may be verified indirectly through the verifications of inflationary cosmological models.

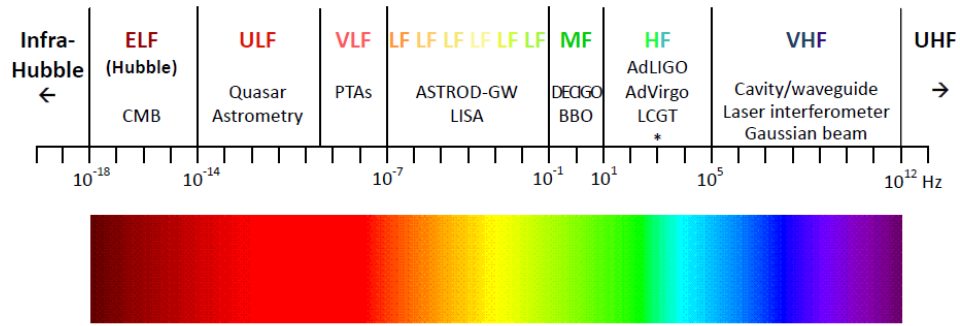

\* AIGO, AURIGA, ET, EXPLORER, GEO, NAUTILUS, MiniGRAIL, Schenberg.

Fig. 2. Frequency classification of GWs.

Fig. 2 shows the frequency classification.[22] For a brief discussion and references of detection methods, please see Ref. [1]. The frequency bands (i) – (vi) are accessible to direct detection, assuming an observation time of up to 100 years (corresponding to observable frequency of about 300 pHz). Fig. 3 gives a comparison of current and planned GW detectors.[23,24] A brief summary of present status can be found in Ref. [25].

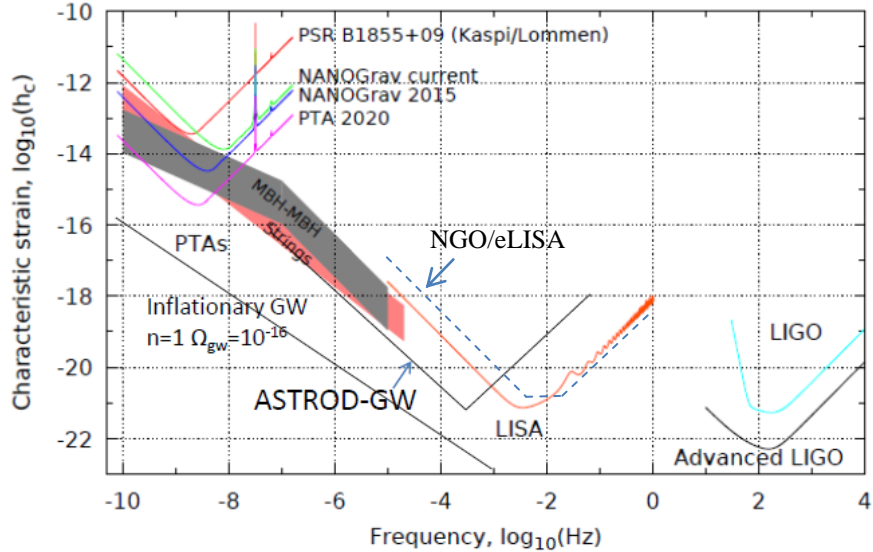

Fig. 3. Comparison of current and planned GW detectors from Demorest *et al.* (2009),[23] showing characteristic strain ($h_c$) sensitivity versus frequency along with expected source strengths. LIGO, LISA and PTAs occupy complementary parts of the GW spectrum. *There is an outstanding gap in the detection band 100 nHz to 10 μHz in the original diagram*. The gray strip is the region all current models of MBH-MBH GW backgrounds occupy. Added to the original diagram are the ASTROD-GW[1,14-16] and NGO/eLISA[18] sensitivity curves and the $\Omega_{gw} = 10^{-16}$ inflationary GW background line. ASTROD-GW has the best sensitivity in the 100 nHz – 1 mHz band. *The outstanding gap in the detection band 100 nHz to 10 μHz is filled by ASTROD-GW. The MBH-MBH GW backgrounds of all current models are above the ASTROD-GW sensitivity level.* The line in the bottom left corner corresponds to $\Omega_{gw} = 10^{-16}$ inflationary GW background ($\Omega_{gw}(f)$ in the figure is decadal density in terms of critical density of the universe defined to be = $(1/\rho_c)$ $(d\rho_{gw}/d\log f)$ with $\rho_{gw}$ the energy density of the stochastic GW background and $\rho_c$ the present value of the critical density for closing the universe in general relativity.).

In Sec.'s 2 and 3, we present the frequency sensitivity spectrum and discuss the scientific goals of ASTROD-GW. In Sec. 4, we present the basic orbit configuration. In Sec. 5, we summarize the CGC ephemeris which is used for orbit simulation and optimization presented in Sec. 6. In Sec.7, we discuss the deployment of ASTROD-GW spacecraft, the propellant ratios and the total mass requirement. In Sec. 8, we discuss time delay interferometry using the CGC ephemeris. In Sec. 9, we discuss the payload. In Sec. 10, we summarize the paper and present an outlook.

## 2. Frequency Sensitivity Spectrum

As shown in Fig. 4, typical frequency sensitivity spectrum of strain for gravitational-wave detection consists of 3 regions, the vibration (seismic noise) or acceleration noise region, the shot noise (flat for current space detector projects) region and the antenna response region. The low-frequency region for the detector sensitivity is dominated by vibration, acceleration noise or gravity-gradient noise. The high-frequency part of the detector sensitivity is restricted by antenna response (or storage time). In a power-limited design, sometimes there is a middle flat region in which the sensitivity is limited by the photon shot noise.[17,20,26]

The shot noise sensitivity limit in the strain for gravitational-wave detection is inversely proportional to $P^{1/2}l$ with $P$ the received power and $l$ the distance. Since $P$ is inversely proportional to $l^2$ and $P^{1/2}l$ is constant, this sensitivity limit is independent of the distance. For 1-2 W emitting power, the limit is $10^{-21}/(Hz)^{1/2}$. As noted in the LISA study,[17] making the arms longer shifts the whole time-integrated sensitivity curve to lower frequencies while leaving the bottom of the curve at the same level. With the same laser power, the ASTROD-GW sensitivity would be shifted to lower frequency by a factor up to 52 if other frequency-dependent requirements can be shifted and met. The sensitivity curve would then be shifted toward lower frequency as a whole. Since the main constraints on the lower frequency part of the sensitivity is from the accelerometer noise, this translational shift depends whether the accelerometer noise requirement for ASTROD-GW could be lowered (more stringent) from that of LISA requirement at a particular frequency. Since ASTROD is in a time frame later than LISA, if the absolute metrological accelerometer/inertial sensor could be developed, there is a potential to go toward this requirement.

However, *to be simple*, here we first take a conservative stand and assume that *the LISA accelerometer noise goal and all other local requirements are taken as they are*. Since the strain sensitivity is mainly the accelerometer noise divided by arm length at low frequency, at a particular low frequency limited by accelerometer noise, the strain sensitivity for ASTROD-GW is 52 times lower than LISA due to longer arm length. With better lower-frequency resolution, the confusion limit for LISA should be somewhat lowered for ASTROD-GW. More studies on confusion limit for ASTROD-GW are much encouraged.

Now we turn to analytic expressions. In the Mock LISA Data Challenge (MLDC) program, the consensus for the LISA instrumental noise density amplitude $^{(MDLC)}S_{Ln}^{1/2}(f)$ is

$$^{(MDLC)}S_{Ln}^{1/2}(f) = (1/L_L) \times \{[(1 + 0.5(f/f_L)^2)] \times S_{Lp} + [1 + (10^{-4}/f)^2](4S_a/(2\pi f)^4)\}^{1/2} \text{Hz}^{-1/2}, \quad (2a)$$

where $L_L = 5 \times 10^9$ m is the LISA arm length, $f_L = c/(2\pi L_L)$ is the LISA arm transfer frequency, $S_{Lp} = 4 \times 10^{-22}$ m$^2$ Hz$^{-1}$ is the LISA (white) position noise level due to photon shot noise, and $S_a = 9 \times 10^{-30}$ m$^2$ s$^{-4}$ Hz$^{-1}$ is the LISA white acceleration noise level.[27] Note that (2a) contains the "reddening" factor $[1 + (10^{-4}/f)^2]$ in the acceleration noise term.

In 2003, Bender[28] looked into the possible LISA sensitivity below 100 μHz. From a careful analysis of noises of test mass and capacitive sensing, Bender suggested a specific sensitivity goal at frequencies down to 3 μHz which contained a milder (than MDLC) "reddening factor". For frequency between 10 μHz to 100 μHz, he suggested to put in the "reddening factor" $[(10^{-4}/f)^{1/2}]$ and for frequency between 3 μHz to 10 μHz, the "reddening factor" $[(10^{-5}/f)]$. To drop this "reddening factor" might be difficult. However, with monitoring the gap of capacitive sensing and the positions of major mass distribution, the factor may be alleviated to certain extent. To completed drop the factor or to go beyond, one may need to go to optical sensing and optical feedback control.[29-37] If we drop the "reddening factor", the LISA instrumental noise density amplitude $^{(Enhanced)}S_{Ln}^{1/2}(f)$ becomes

$$^{(Enhanced)}S_{Ln}^{1/2}(f) = (1/L_L) \times \{[(1 + 0.5(f/f_L)^2)] \times S_{Lp} + [4S_a/(2\pi f)^4]\}^{1/2} \text{ Hz}^{-1/2}. \quad (2b)$$

For ASTROD-GW, our goal on the instrumental strain noise density amplitude is

$$S_{An}^{1/2}(f) = (1/L_A) \times \{[(1 + 0.5(f/f_A)^2)] \times S_{Ap} + [4S_a/(2\pi f)^4]\}^{1/2} \text{ Hz}^{-1/2}, \quad (3)$$

over the frequency range of 100 nHz $< f <$ 1 Hz with $f$ in unit of Hz. Here $L_A = 260 \times 10^9$ m is the ASTROD-GW arm length, $f_A = c/(2\pi L_A)$ is the ASTROD-GW arm transfer frequency, $S_a = 9 \times 10^{-30}$ m$^2$ s$^{-4}$ Hz$^{-1}$ is the white acceleration noise level (the same as that for LISA), and $S_{Ap} = 10816 \times 10^{-22}$ m$^2$ Hz$^{-1}$ is the (white) position noise level due to laser shot noise $52^2$ times that for LISA).[1,14-16,24]

The corresponding noise curve for the ASTROD-GW instrumental noise density amplitude $^{(MDLC)}S_{An}^{1/2}(f)$ with the same "reddening" factor as specified in MLDC program is

$$S_{An}^{1/2}(f) = (1/L_A) \times \{[(1 + 0.5(f/f_A)^2)] \times S_{Ap} + [1 + (10^{-4}/f)^2](4S_a/(2\pi f)^4)\}^{1/2} \text{ Hz}^{-1/2}, \quad (3a)$$

over the frequency range of 100 nHz $< f <$ 1 Hz with $f$ in unit of Hz. The sensitivity curves from the four formulas (2a), (2b), (3) and (3a) are shown in Fig. 4 for easy comparison.

Since the arm length is longer than LISA by 52 times, with 1-2 W laser power and LISA acceleration noise, the strain sensitivity of ASTROD-GW is 52 times lower than LISA in the lower frequency region whether we take (2a) and (3a) to compare or (2b) and (3) to compare, and is better than LISA and Pulsar Timing Arrays (PTAs) in the frequency band 100 nHz – 1 mHz (Fig. 3). ASTROD-GW will complement LISA and PTAs in exploring single events and backgrounds of MBH-MBH binary GWs in the important frequency range 100 nHz - 1 mHz to study black hole co-evolution with galaxies, dark energy and other issues (Sec. 3).

Although how well one can improve on the lower frequency part of LISA and NGO/eLISA accelerometer/inertial sensor noise over the MLDC formula is not clear, *the strain sensitivity of ASTROD-GW in the lower frequency part is better than LISA and NGO/eLISA by 52 and 260 times respectively due to longer arm length if one assumes* *the same accelerometer/inertial sensor noise*. For ASTROD-GW, we consider Eq. (3) as our goal and Eq. (3a) as our requirement of instrumental strain noise density amplitude.

After NASA's withdrawal from ESA-NASA collaboration of LISA in 2011, the European NGO/eLISA for space detection of GWs emerged. The orbit configuration is the same as LISA, but with arm length shrunk 5 times to one million kilometers, the orbits slowly drifting away from the Earth and the nominal mission duration 2 years (extendable to 5 years) to save weight, fuel and costs. The three spacecraft will consist of one "mother" and two simpler "daughters," with interferometric measurements along only two arms with the "mother" at the vertex.[18] The NGO/eLISA strain noise power-spectral-density goal[18,38] is also shown in Fig. 4. For the lower frequency part of the power spectrum of NGO/eLISA, we choose to use the same acceleration noise as enhanced version of LISA to obtain the NGO/eLISA strain noise for easy comparison.

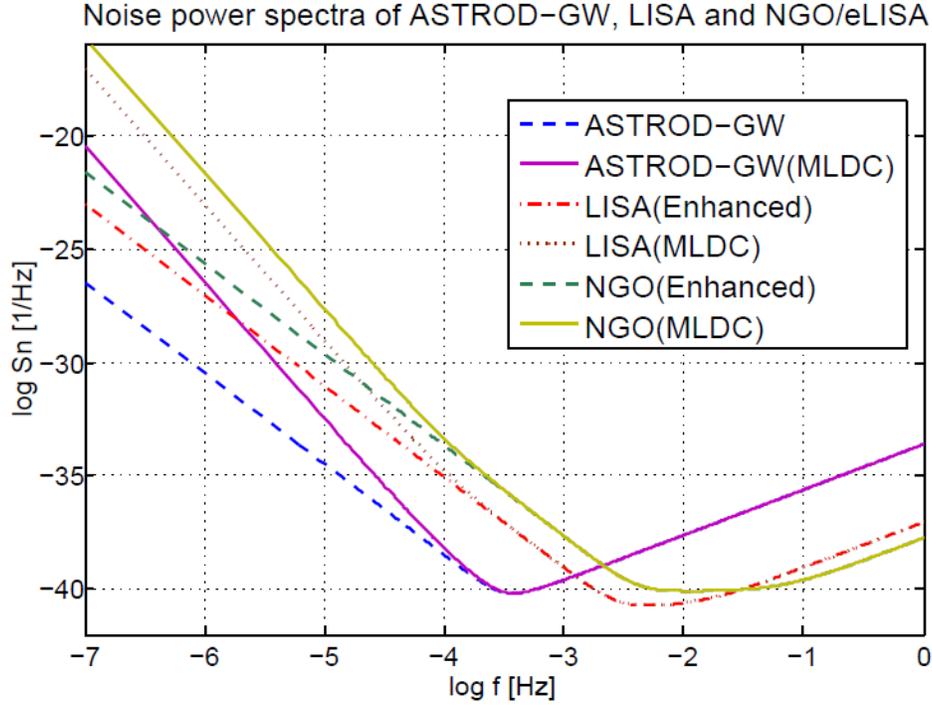

Fig. 4. Strain noise power spectra of ASTROD-GW as compared with LISA and NGO/eLISA.

## 3. Scientific Goals

In this section, we review and summarize the scientific goals for the ASTROD-GW mission project.[1,14-16,24] More studies on scientific goals are needed for ASTROD-GW in the next few years

### 3.1. Massive Black Holes (MBHs) and their co-evolution with galaxies

Observational evidence indicate that massive black holes (MBHs) reside in most local galaxies. Relations have been discovered between the MBH mass and the mass of host galaxy bulge, and between the MBH mass and the velocity-dispersion. These relations indicate that the central MBHs are linked to the evolution of galactic structure. Newly fueled quasar may come from the gas-rich major merger of two massive galaxies. Recent astrophysical evidence links together these major galaxy mergers and the growth of supermassive black holes in quasars.[39,40] Distant quasar observations indicate that MBH of billions of solar masses already existed less than a billion years after the Big Bang. At present, there are different theoretical proposals for scenarios of the initial conditions and formations of black holes. These scenarios include BH seeds from inflationary Universe and/or from the collapse of Population III stars, different accretion models and binary formation rates. All of these models generate MBH merging scenarios in galaxy co-evolution with GW radiations. Measurement of amplitude and spectrum of these GWs will tell us the MBH cosmic history.

The standard theory of massive black hole formation is the merger-tree theory with various Massive Black Hole Binary (MBHB) inspirals acting. The GWs from these MBHB inspirals can be detected and explored to cosmological distances using space GW detectors. Although there are different merger-tree models and models with BH seeds, they all give significant detection rates for space GW detectors and Pulsar Timing Arrays (PTAs).[41,42,23] Gravitational wave (GW) observation in the 300 pHz – 0.1 Hz frequency band will be a major observation tool to study the co-evolution of galaxy with BHs. This frequency band covers the low frequency band (100 nHz - 100 mHz) and very low frequency band (300 pHz-100 nHz) GWs and is in the detection range of PTAs,[43,44] NGO/eLISA and ASTROD-GW. PTAs are most sensitive in the frequency range 300 pHz -100nHz, NGO/eLISA space GW detector is most sensitive in the frequency range 2 mHz – 0.1 Hz, while ASTROD-GW is most sensitive in the frequency range 100 nHz - 2 mHz (Fig. 3). PTAs have already been collecting data for detection of stochastic GW background from MBH binary mergers, and are aiming at detection around 2020.[23] NGO/eLISA and ASTROD-GW will be able to directly observe how massive black holes form, grow, and interact over the entire history of galaxy formation. ASTROD-GW will detect stochastic GW background from MBH binary mergers in the frequency range 100 nHz to 100 μHz. These observations are significant and important to the study of co-evolution of galaxies with MBHs. The expected rate of MBHB sources is 10 yr$^{-1}$ to 100 yr$^{-1}$ for NGO/eLISA and 10 yr$^{-1}$ to 1000 yr$^{-1}$ for LISA.[18] For ASTROD-GW, we should expect similar number of sources but with better angular resolution (Sec. 4.3).

At present, there are different theoretical scenarios for the initial conditions and formations of black holes, e.g., primordial massive BH clouds as seeds, direct formation of supermassive black hole via multi-scale gas inflows in galaxy mergers, direct collapse into a supermassive black hole from mergers between massive protogalaxies with no need to suppress cooling and star formation, etc. The mass range and maximum mass of Population III stars is also a relevant issue for seed BHs. A strip in Fig. 3 covers possible scenarios. ASTROD-GW with sensitivity below this strip will be able to detect GW background for every model in this region and to distinguish various scenarios for finding the history of BH and galaxy co-evolution.

With the detection of MBH merger events and background, the properties and distribution of MBHs could be deduced and underlying population models could be tested. Sesana et al.[45] consider and compare ten specific models of massive black hole formation. These models are chosen to probe four important and largely unconstrained aspects of input physics used in the structure formation simulations, i.e., seed formation, metallicity feedback, accretion efficiency and accretion geometry. With Bayesian analyses to recover posterior probability distribution, they show that LISA has enormous potential to probe the underlying physics of structure formation. With better sensitivity in the frequency range 100 nHz - 1 mHz, ASTROD-GW will be able to probe the underlying physics of structure formation further. With the detection of the GW background of the MBH mergers, PTAs and ASTROD-GW will add to our understanding of the structure formation.

### 3.2. Extreme Mass Ratio Inspirals (EMRIs)

EMRIs are GW sources for space GW detectors. The NGO/eLISA sensitive range for central MBH masses is $10^4$-$10^7$ $M_\odot$. The expected number of NGO/eLISA detections over two years is 10 to 20;[18] for LISA, a few tens;[18] for ASTROD-GW, similar or more with sensitivity toward larger central BH's and with better angular resolution (Sec. 4.3).

### 3.3. Testing Relativistic Gravity

An important scientific goal of LISA[17] and NGO/eLISA[18,28] is to test general relativity and to study black hole physics with precision in strong gravity. With better precision in 100 nHz-1 mHz frequency range, ASTROD is going to push this goal further in many aspects. These include testing strong-field gravity, precision probing of Kerr spacetime and measuring/constraining the mass of graviton. Some considerations have been given in Ref.s 46, 47. Lower frequency sensitivity is significant in improving the precision of various tests.[46,47] Further studies in these respects would be of great value.

### 3.4. Dark energy and cosmology

In the dark energy issue,[48] it is important to determine the value of $w$ in the equation of state of dark energy,

$$w = p / \rho, \qquad (4)$$

as a function of different epochs where $p$ is the pressure and $\rho$ the density of dark energy. For cosmological constant as dark energy, $w = -1$. From cosmological observations, our universe is close to being flat. In a flat Friedman Lemaître-Robertson-Walker (FLRW) universe, the luminosity distance is given by

$$d_L(z) = (1+z) \int_{0 \to z} (H_0)^{-1} [\Omega_m(1+z')^3 + \Omega_{DE}(1+z')^{3(1+w)}]^{-(1/2)} dz', \qquad (5)$$

where $H_0$ is Hubble constant, $\Omega_{DE}$ is the present dark energy density parameter, and the equation of state of the dark energy $w$ is assumed to be constant. In the case of non-constant $w$ and non-flat FLRW universe, similar but more complicated expression can be derived. Here we show (5) for illustrative purpose. From the observed relation of luminosity distance vs. redshift $z$, the parameter $w$ of the equation of state as a function of redshift $z$ can be solved for and compared with various cosmological models. Dark energy cosmological models can be tested this way. Luminosity distance from supernova observations and from gamma ray burst observations vs. redshift observations are the focus for the current dark energy probes.

Space GW detectors observing MBHB inspirals and EMRIs are good probes to determine the luminosity distances. With the redshift of the source determined by the electromagnetic observations of associated galaxies or cluster of galaxies, these space GW detectors are also dark energy probes. In the merging of MBHs during the galaxy co-evolution processes, gravitational waveforms generated give precise, gravitationally calibrated luminosity distances to high redshift. The inspiral signals of these binaries can serve as standard candles/sirens.[49,50] With better angular resolution, ASTROD-GW will have better chance to identify the associated electromagnetic redshift and therefore will be better for the determination of the dark energy equation of state (Sec. 4.3).[24]

### 3.5. Compact binaries

Space GW detectors are also sensitive to the GWs from Galactic compact binaries.[17,18] These detectors will be able to survey compact stellar-mass binaries and study the structure of the Galaxy. NGO/eLISA will detect about 3000 double white dwarf binaries individually with most in the GW frequency band 3-6 mHz (orbit period about 300-600 s); for LISA, about 10,000 double white dwarf binaries. These sources constitute the population which has been proposed as progenitors of normal type Ia and peculiar supernovae. For a review on the electromagnetic counterparts of GW mergers of compact objects, see, e.g., Ref. 51. At the frequency band 3-6 mHz, NGO/eLISA is more sensitive than ASTROD-GW (Fig. 4). Since NGO/eLISA will be flying first these GW signals will serve as a calibration for ASTROD-GW in addition to the verification binaries.

At GW frequencies below a few mHz, millions of ultra-compact binaries will form a detectable foreground for NGO/eLISA and ASTROD-GW. At these frequencies, ASTROD-GW is more sensitive than NGO/eLISA (Fig. 4). More sources will be resolved individually and ASTROD-GW can improve on the observational results of NGO/eLISA.

### 3.6. Relic GWs

The straight line in the bottom left corner of Figure 3 corresponds to $\Omega_{gw} = 10^{-16}$ inflationary GW background. For ASTROD-GW, when a 6-S/C formation is used for correlated detection of stochastic GWs, the sensitivity can reach this region. However, the anticipated MBH-MBH GW background is above the 3-S/C ASTROD-GW sensitivity. If this background is detected, then the detectability of inflationary gravitational wave of the strength $\Omega_{gw}$ = $10^{-16}$-$10^{-17}$ from 6-S/C formation in the ASTROD-GW frequency region depends on whether this MBH-MBH GW 'foreground' could be separated due to different frequency dependence or other signatures.

For direct detection of primordial (inflationary, relic) GWs in space, one may go to frequencies lower or higher than LISA bandwidth,[1,52] where there are potentially less foreground astrophysical sources[53] to mask detection. DECIGO[54,55] and Big Bang Observer[56] look for GWs in the higher frequency range while ASTROD-GW looks for GWs in the lower frequency range. Their instrument sensitivity goals all reach $10^{-17}$ in terms of critical density. The main issue is the level of foreground and whether foreground could be separated.

Other potentially possible GW sources in the relevant frequency band, e.g., cosmic strings, should also be studied.

## 4. Basic Orbit Configuration

### 4.1. Basic concept

The basic ASTROD-GW configuration consists of three spacecraft in the vicinity of the Sun-Earth Lagrange points L3, L4 and L5 respectively with near-circular orbits around the Sun, forming a nearly equilateral triangle as shown by Fig. 1 with the three arm lengths about $2.6 \times 10^8$ km (1.732 AU). The dominant force on the spacecraft is from the Sun in the restricted three-body problem of Earth-Sun-spacecraft system. Since the Earth-Sun orbit is elliptical, the Lagrange points are not stationary in the Earth-Sun rotating frame. The motion of test particles at L3, L4 and L5 deviates from circular orbit by a fraction of $O(e)$ where $e$ (=0.0167) is the eccentricity of the Earth orbit around the Sun. However, the spacecraft can be in the halo orbit of the respective Lagrange points largely compensating the non-stationary motion of the Lagrange points to remain nearly circular orbits of the Sun. The circular orbits of spacecraft near the L3, L4 and L5 points are stable or virtually stable in 20 years (their orbits are also stable or quasi-stable with respect to their respective Lagrange points so that the deviations from circular orbit of their respective Lagrange point are of the order of $O(e^2)$ in AU ) and the deviation of the spacecraft triangle from an equilateral triangle is of order of $O(e^2)$ in arm length. When the orbits of spacecraft have a small inclination $\lambda$ (in radians) with respect to the ecliptic plane the arm length variation is of the order of $O(\lambda^2)$. Therefore the added variation due to these two causes is of the order $O(e^2, \lambda^2)$. For these two causes to match (to $O(10^{-4})$), $\lambda$ should be of the order of $O(1°)$. Later in Sec. 6, we will see that the perturbation from all planets except Earth is of the order of $O(10^{-4})$. The influence of Earth is already taken into consideration since the L3, L4 and L5 points are effectively stable in 20 years. Hence, suitable inclined circular orbits could be our basic orbits to start with and the deviations from actual optimized orbit should be on the order of $O(10^{-4})$.

### 4.2. Basic orbit design with inclination

In the original proposal, the ASTROD-GW orbits are chosen with $\lambda = 0$ in the ecliptic plane. The angular resolution in the sky has antipodal ambiguity and, near ecliptic poles, the resolution is poor, although over most of sky the resolution is good. Now we have designed the basic orbits of ASTROD-GW to have small inclinations to resolve these issues while keeping the variation of the arm lengths in the tolerable range.[24]

The basic idea is that if the orbits of the ASTROD-GW spacecraft are inclined with a small angle λ, the interferometry plane with appropriate design is also inclined with similar angle and when the ASTROD-GW formation evolves, the interferometry plane modulates in the ecliptic solar-system barycentric frame. With this, angular positions of GW sources both near the polar region and off the polar region are resolved without antipodal ambiguity.

Let first consider a circular orbit of a spacecraft in the Newtonian gravitational central problem (one-body) in spherical coordinates $(r, \theta, \varphi)$:

$$r = a, \theta = 90°, \varphi = \omega t + \varphi_0, \tag{6}$$

where $a$, $\omega$, and $\varphi_0$ are constants. For spacecraft in this discussion, we have $a = 1$ AU, $\omega = 2\pi/T_0$ with $T_0 = 1$ sidereal year, and $\varphi_0$ is the initial phase in the coordinate considered. The spacecraft orbit at time $t$ in Cartesian coordinates is

$$x = a \cos \varphi = a \cos (\omega t + \varphi_0); y = a \sin \varphi = a \sin (\omega t + \varphi_0); z = 0. \tag{7}$$

Let transform this orbit actively into an orbit with inclination, and with the intersection of the orbit plane and *xy*-plane at the line $\varphi = \Phi_0$ in the *xy*-plane. The active transformation matrix is

$$R(\lambda; \Phi_0) = \begin{bmatrix} \cos^2\Phi_0 + \sin^2\Phi_0\cos\lambda & \sin\Phi_0\cos\Phi_0(1 - \cos\lambda) & \sin\Phi_0\sin\lambda \\ \sin\Phi_0\cos\Phi_0(1 - \cos\lambda) & \sin^2\Phi_0 + \cos^2\Phi_0\cos\lambda & -\cos\Phi_0\sin\lambda \\ -\sin\Phi_0\sin\lambda & \cos\Phi_0\sin\lambda & \cos\lambda \end{bmatrix}. \tag{8}$$

The new spacecraft orbit is

$$\begin{bmatrix} x' \\ y' \\ z' \end{bmatrix} = \begin{bmatrix} a[1 - \sin^2\Phi_0(1 - \cos\lambda)]\cos\varphi + a \sin\Phi_0\cos\Phi_0(1 - \cos\lambda)\sin\varphi \\ a \cos\Phi_0\sin\Phi_0(1 - \cos\lambda)\cos\varphi + a[1 - \cos^2\Phi_0(1 - \cos\lambda)]\sin\varphi \\ -a \sin\Phi_0\sin\lambda\cos\varphi + a \cos\Phi_0\sin\lambda\sin\varphi \end{bmatrix}. \tag{9}$$

For the three orbits with inclination $\lambda$ (in radian), we choose:

S/C I: $\Phi_0(I) = 270°$, $\varphi_0(I) = 0°$;
S/C II: $\Phi_0(II) = 150°$, $\varphi_0(II) = 120°$;
S/C III: $\Phi_0(III) = 30°$, $\varphi_0(III) = 240°$. (10)

Defining

$$\xi = 1 - \cos \lambda = 0.5 \lambda^2 + O(\lambda^4), \tag{11}$$

from Eq. (9) and Eq. (10), we have

(i) for the orbit of S/C I

$$\begin{bmatrix} x^I \\ y^I \\ z^I \end{bmatrix} = \begin{bmatrix} a \cos \omega t - \xi a \cos \omega t \\ a \sin \omega t \\ a \cos \omega t \sin \lambda \end{bmatrix}, \quad (12)$$

(ii) for the orbit of S/C II

$$\begin{bmatrix} x^{II} \\ y^{II} \\ z^{II} \end{bmatrix} = \begin{bmatrix} a[(-\tfrac{1}{2})\cos \omega t - (3^{1/2}/2) \sin \omega t] + (a/2) \xi[(3^{1/2}/2) \sin \omega t - \tfrac{1}{2} \cos \omega t] \\ a[(-\tfrac{1}{2})\sin \omega t + (3^{1/2}/2) \cos \omega t] + (3^{1/2}/2) a \xi[(3^{1/2}/2) \sin \omega t - \tfrac{1}{2} \cos \omega t] \\ a \sin \lambda\, [(3^{1/2}/2) \sin \omega t - \tfrac{1}{2} \cos \omega t] \end{bmatrix}, \quad (13)$$

(iii) for the orbit of S/C III.

$$\begin{bmatrix} x^{III} \\ y^{III} \\ z^{III} \end{bmatrix} = \begin{bmatrix} a[(-\tfrac{1}{2})\cos \omega t + (3^{1/2}/2) \sin \omega t] + (a/2) \xi[(3^{1/2}/2) \sin \omega t - \tfrac{1}{2} \cos \omega t] \\ a[(-\tfrac{1}{2})\sin \omega t - (3^{1/2}/2) \cos \omega t] - (3^{1/2}/2) a \xi[(-3^{1/2}/2) \sin \omega t - \tfrac{1}{2} \cos \omega t] \\ a \sin \lambda\, [(-3^{1/2}/2) \sin \omega t - \tfrac{1}{2} \cos \omega t] \end{bmatrix}. \quad (14)$$

It is good to check that $[(x^I)^2 + (y^I)^2 + (z^I)^2]^{1/2} = [(x^{II})^2 + (y^{II})^2 + (z^{II})^2]^{1/2} = [(x^{III})^2 + (y^{III})^2 + (z^{III})^2]^{1/2} = a$ hold for consistency.

We can now calculate the arm vectors $\underline{V}_{II-I} = \underline{r}^{II} - \underline{r}^I$, $\underline{V}_{III-II} = \underline{r}^{III} - \underline{r}^{II}$ and $\underline{V}_{I-III} = \underline{r}^{III} - \underline{r}^I$:

$$\underline{V}_{II-I} = \begin{bmatrix} a[-(3/2) \cos \omega t - (3^{1/2}/2) \sin \omega t] + a \xi[(3^{1/2}/4) \sin \omega t + (3/4) \cos \omega t] \\ a[-(3/2) \sin \omega t + (3^{1/2}/2) \cos \omega t] + a \xi[(3/4) \sin \omega t - (3^{1/2}/4) \cos \omega t] \\ a \sin \lambda\, [(3^{1/2}/2) \sin \omega t - (3/2) \cos \omega t] \end{bmatrix}, \quad (15)$$

$$\underline{V}_{III-II} = \begin{bmatrix} 3^{1/2} a \sin \omega t - (3^{1/2}/2) a \xi \sin \omega t \\ -3^{1/2} a \cos \omega t + (3^{1/2}/2) a \xi \cos \omega t \\ -3^{1/2} a \sin \lambda \sin \omega t \end{bmatrix}, \quad (16)$$

$$\underline{V}_{I-III} = \begin{bmatrix} a[(3/2) \cos \omega t - (3^{1/2}/2) \sin \omega t] + a \xi[(3^{1/2}/4) \sin \omega t - (3/4) \cos \omega t] \\ a[(3/2) \sin \omega t + (3^{1/2}/2) \cos \omega t] + a \xi[-(3/4) \sin \omega t - (3^{1/2}/4) \cos \omega t] \\ a \sin \lambda\, [(3^{1/2}/2) \sin \omega t + (3/2) \cos \omega t] \end{bmatrix}, \quad (17)$$

The closure relation $\underline{V}_{II-I} + \underline{V}_{III-II} + \underline{V}_{I-III} = \underline{0}$ is checked for verifying calculations. The arm lengths are calculated to be

$$|\underline{V}_{II-I}| = 3^{1/2} a\, [(1 - \xi/2)^2 + \sin^2 \lambda \sin^2 (\omega t - 60°)]^{1/2},$$

$$|\underline{V}_{III-II}| = 3^{1/2} a\, [(1 - \xi/2)^2 + \sin^2 \lambda \sin^2 (\omega t)]^{1/2},$$

$$|\underline{V}_{I-III}| = 3^{1/2} a\, [(1 - \xi/2)^2 + \sin^2 \lambda \sin^2 (\omega t + 60°)]^{1/2}. \quad (18)$$

The fractional arm length variation is within $(1/2) \sin^2 \lambda$ which is about $10^{-4}$ for $\lambda$ about $1°$.

The cross-product vector $\underline{N}(t) \equiv \underline{V}_{III-II} \times \underline{V}_{I-III}$ is normal to the orbit configuration plane and has the following components:

$$\underline{N} = [(3^{3/2}/2)\,(1 - \xi/2)\,a^2] \begin{bmatrix} -\sin\lambda\cos 2\omega t \\ -\sin\lambda\sin 2\omega t \\ (1 - \xi/2) \end{bmatrix}. \quad (19)$$

The normalized unit normal vector $\underline{n}$ is then:

$$\underline{n} = [\sin^2\lambda + (1 - \xi/2)^2]^{1/2} \begin{bmatrix} -\sin\lambda\cos 2\omega t \\ -\sin\lambda\sin 2\omega t \\ (1 - \xi/2) \end{bmatrix}. \quad (20)$$

The geometric center $\underline{V}_c$ of the ASTROD-GW spacecraft configuration is

$$\underline{V}_c = \begin{bmatrix} -(1/2)\,\xi\,a\cos\omega t \\ (1/2)\,\xi\,a\sin\omega t \\ 0 \end{bmatrix}. \quad (21)$$

There are 3 interferometers with 2 arms in the ASTROD-GW configuration. The geometric center of each of these 3 interferometers is at a distance of about 0.25 AU from the Sun. Numerical simulation and optimization of orbit configuration for inclination of 1° is worked out in Ref. 57 and summarized in Sec. 6.2 using planetary ephemeris to take into account of the planetary perturbations. For inclinations between 1° to 3°, optimizations are underway. When LISA orbits around the Sun, it is equivalent to multiple detector arrays distributed in 1 AU orbit. The extension of ASTROD-GW is already of 1 AU. When ASTROD-GW orbits around the Sun, it is also equivalent to multiple detector arrays distributed in 1 AU orbit.

### 4.3. Angular resolution

Now let us consider angular resolution of a coherent GW source. Consider first the LISA case as example. The detector formation of LISA is modulated in its orbit around the Sun. The azimuth modulation amplitude is $2\pi$ rad with inclination 1.05 rad (60°) so that the antenna pattern sweeps around the sky in one year. The antenna response is not isotropic but the averaged linear angular resolution (in a year) of monochromatic GW sources for LISA differs by less than a factor of 3 among all directions. *The angular resolution is basically proportional to the inverse of the strain signal to noise ratio.* If the inclination is of the order 0.017-0.052 rad (1-3°) for LISA, the polar resolution would be worsened by 30-10 times (approximately the ratio sin of 1.05 rad to sin of 0.03-0.1 rad). The steradian localization in the celestial sphere is worsened by square of this factor; away from the polar region ($\theta \gg 0.017$-$0.052$ rad) by $\sin^2\theta$. If the signal to noise ratio is downgraded by 5 (as in eLISA/NGO in its low frequency part due to shorter arm length), the linear angular resolution is worsened by 5 times. ASTROD-GW has less sensitivity above 1 mHz compared with LISA, therefore the angular resolution will be worsened by both factors. In the 100 nHz-1 mHz region, ASTROD-GW has better sensitivity compared with LISA, in most part by 52 times. Hence, the angular resolution in the polar region is similar to that of LISA, while in other regions, the linear resolution is enhanced by roughly 52 × sin $\theta$ (upgraded by 52 but downgraded by $\sin^2 \theta$ in sterad [by sin $\theta$ in rad]). Although there is a mild dependence on the configuration inclination angle $\lambda$, within a factor of 3, the averaged antenna pattern for ASTROD-GW away from the polar region is better by a factor of 52 × sin $\theta$ compared to that of LISA. Since the antenna pattern of ASTROD-GW sweeps over the whole sky in half year as can be seen from Eq. (20), the time of average needed is half a year instead of a year.[24]

For more complicated sources like chirping GW sources from BBHs (binary black holes), one needs to do fitting in order to obtain the accuracy of the parameters. However, the tendency of accuracy of parameters is the same: for similar situations, it is proportional to the inverse of the strain signal to noise ratio.

### 4.4. Six spacecraft formation

For a more sensitive detection of background or relic GWs, correlated detection with 2 sets of triangular ASTROD-GW formation are required, i.e., a 6-S/C constellation. Since the triangular formation with 3 S/C near L3, L4, L5 of Sun-Earth system respectively is unique in orbit stability, to place the second triangular formation shifted by a few degree or more with respect to the first one in 1 AU orbit around the Sun is not desirable in stability. However, we could put the second triangular formation again near L3, L4, L5 respectively, but separated from the first formation by $1 \times 10^6$ km to $5 \times 10^6$ km for the respective S/C.

## 5. CGC Ephemeris

In 1998, we started orbit simulation and parameter determination for ASTROD.[58,59] We worked out a post-Newtonian ephemeris of the Sun including the solar quadrupole moment, the eight planets, the Pluto, the moon and the 3 biggest asteroids. We term this working ephemeris CGC 1 (CGC: Center for Gravitation and Cosmology). Using this ephemeris as a deterministic model and adding stochastic terms to simulate noise, we generate simulated ranging data and use Kalman filtering to determine the accuracies of fitted relativistic and solar-system parameters for 1050 days of the ASTROD mission.

For a better evaluation of the accuracy of $\dot{G}/G$, we need also to monitor the masses of other asteroids. For this, we considered all known 492 asteroids with diameter greater than 65 km to obtain an improved ephemeris framework --- CGC 2, and calculated the perturbations due to these 492 asteroids on the ASTROD spacecraft.[60,61]

In building CGC ephemeris framework, we use the post-Newtonian barycentric metric and equations of motion as derived in Brumberg[62] with PPN (Parametrized Post-Newtonian) parameters $\beta$ and $\gamma$ for solar system bodies (with the gauge parameter α set to zero). These equations are used to build our computer-integrated ephemeris (with the PPN parameters $\gamma = \beta = 1$, $J_2 = 2 \times 10^{-7}$) for eight-planets, the Pluto, the Moon and the Sun. The positions and velocities at the epoch 2005.6.10 0:00 are taken from the DE403 ephemeris. The evolution is solved by using the 4$^{\text{th}}$-order Runge-Kutta method with the step size h =0.01 day. In Ref. 58, the 11-body evolution is extended to 14-body to include the 3 big asteroids — Ceres, Pallas and Vesta (CGC 1 ephemeris). Since the tilt of the axis of the solar quadrupole moment to the perpendicular of the elliptical plane is small (7°), in CGC 1 ephemeris, we have neglected this tilt. In CGC 2 ephemeris, we have added the perturbations of additional 489 asteroids.

In our first optimization of ASTROD-GW orbits,[63-65] we have used CGC 2.5 ephemeris in which only 3 biggest minor planets are taken into accounts, but the Earth's precession and nutation are added; the solar quadratic zonal harmonic and the Earth's quadratic to quartic zonal harmonic are considered. In later simulation, we add the perturbation of additional 349 asteroids and call it CGC 2.7 ephemeris. The differences in orbit evolution compared with DE405 for Earth for 3700 days starting at JD2461944.0 (2028-Jun-21 12:00:00) are shown in Fig. 5. The differences in radial distances are less than about 200 m. The differences for other inner planets are smaller.

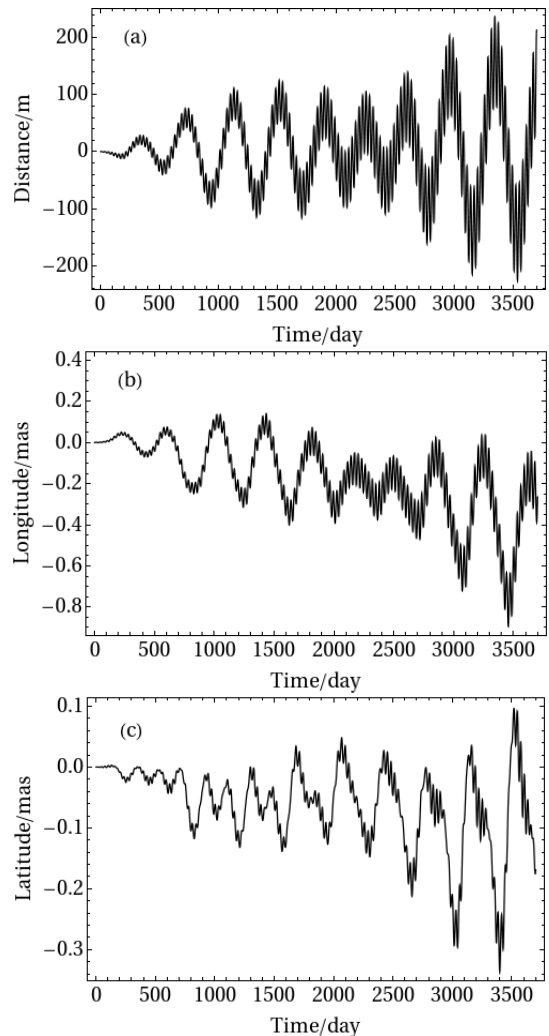

Fig. 5. Differences (DE405-CGC2.7) in (a) radial distance, (b) longitude, (c) latitude starting at JD2461944.0 for Earth orbits in DE405 and CGC2.7 ephemerides in the J2000 Heliocentric Earth mean equatorial coordinate system.

## 6. Orbit Optimization

The goal of the ASTROD-GW mission orbit optimization is to equalize the three arm lengths of the ASTROD-GW formation and to reduce the relative line-of-sight velocities between three pairs of spacecraft as much as possible. In our first optimization, the time of start of the science part of the mission is chosen to be noon, June 21, 2025 (JD2460848.0) and the optimization is for a period of 3700 days using CGC 2.5 ephemeris.[63-65] Since the preparation of the mission may take longer time and there is a potential that the extended mission life time may be longer than 10 years, in a more recent optimization, we start at noon, June 21, 2028 (JD2461944.0) and optimize for a period of 20 years using CGC 2.7 ephemeris including more asteroids.[66-68] In both of these optimizations, the orbit configuration is set in the ecliptic plane and we have the inclination angle $\lambda = 0$. Very recently, we optimize an inclined orbit configuration with $\lambda = 1°$ starting at noon, June 21, 2025 (JD2460848.0) using Newtonian ephemeris.[57] In the following, we present the results of the two more recent optimizations.

### 6.1. Inclination $\lambda = 0$ case

6.1.1. *First choice of initial conditions*

To start the optimization, the initial positions of the three spacecraft are chosen to form an equilateral triangle on 1 AU solar orbit and the initial velocity chosen to follow an orbit period of one sidereal year. To fix the orbit, we tentatively take 12:00 21st June 2028 (JD2461944.0) as the initial time that ASTRO-GW configuration enters the science phase. At this moment the earth is located near the summer solstice, the initial choice of initial positions of spacecraft when they enter into the orbit are predetermined to be the places close to the Lagrange points L3, L4 and L5 at noon, June 21, 2028 (JD2461944.0), in the heliocentric ecliptic coordinate system as follows: S/C 1 (0, 1 AU, 0), S/C 2 ($3^{1/2}/2$ AU, -1/2 AU, 0), and S/C 3 ($-3^{1/2}/2$ AU, -1/2 AU, 0). The initial velocities of the three spacecraft in the heliocentric ecliptic coordinate system are chosen to be

S/C 1 at ($-v_0$, 0, 0), S/C 2 at (0.5 $v_0$, $3^{1/2}2^{-1}$ $v_0$, 0), S/C 3 at (0.5 $v_0$, $-3^{1/2}2^{-1}$ $v_0$, 0),

where $v_0$ (= 0.01720209895 AU/day) is the circular velocity of spacecraft at 1 AU from the Sun. *Note that in this section and in the following two sections* (Sec.'s 6-8)*, the labeling of spacecraft and arms is different from that of* Fig. 1*: we have the correspondence* S/C 1 → S/C 3(Fig. 1), S/C 2 → S/C 2(Fig. 1), S/C 3 → S/C 1(Fig.1), *and* Arm 1 → Arm 1(Fig.1), Arm 2 → Arm 3(Fig.1), Arm 3 → Arm 2(Fig.1).

6.1.2 *Method of optimization*

In the solar system the ASTROD-GW spacecraft orbits are perturbed by the other planets. The largest are due to Jupiter and Venus. Therefore the orbital period and the eccentricity change. Our method of optimization is to modify the initial velocities and initial heliocentric distances so that the perturbed orbital periods for ten or twenty year average remain close to one sidereal year (trimming the period) and that the average eccentricities remain near zero (trimming the eccentricity). The methods of trimming the period and trimming the eccentricity are explained in detail in Ref's 64-68.

At the beginning of an optimization step, we calculate the orbits of the spacecraft from initial conditions using the ephemeris CGC 2.7. From these orbits, we calculate the mean orbital period in 20 years, the variations of arm lengths, differences of arm lengths, heliocentric distances of spacecraft and Doppler velocities between spacecraft as functions of the mission time.

With these data, we optimize the orbital period of each spacecraft. (i) We first calculate the average periods of each spacecraft and change the initial conditions to match the periods to one sidereal year to minimize the change of arm lengths. (ii) Due to planetary perturbations, for the spacecraft near L3 point, the period decreases slightly as time goes on, while for the spacecraft near L4 and L5 points, the period increases slightly as time goes on; Arm 3 changes from short to longer, Arm 2 changes from getting longer to getting shorter, and the Arm 1 changes are relatively small. In this situation, we decrease the initial velocity of S/C 1 to let the initial period slightly longer; and we increase the initial velocities of S/C 2 and S/C 3 to let their initial period slightly shorter. This way, we can correct the period deviation tendency of the three spacecraft, and minimize the period differences and arm length differences dynamically. After trimming the period, the arm length differences are decreased; however, in general, the orbit configuration still does not satisfy the mission requirements.

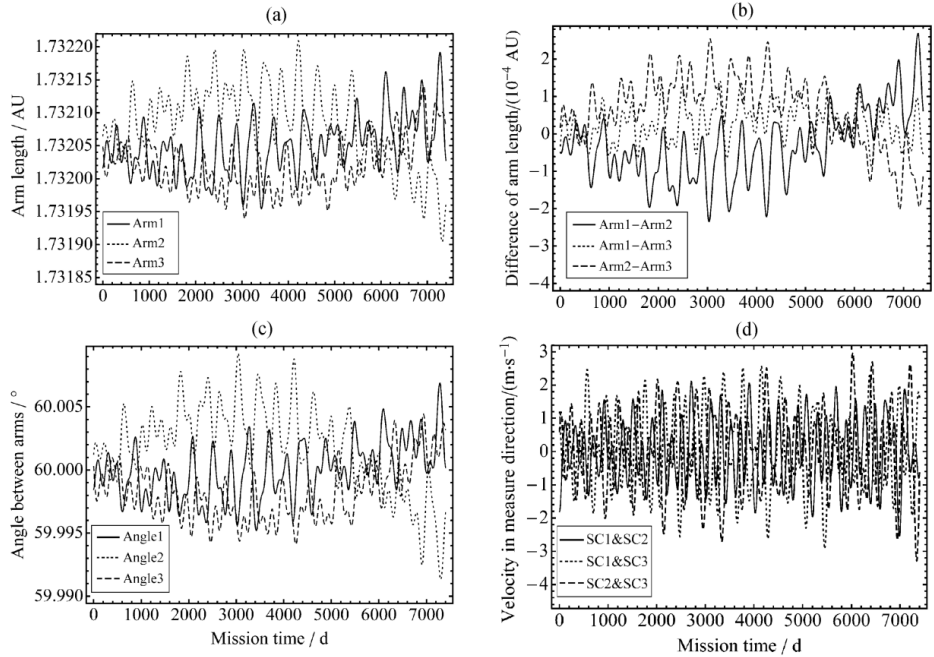

Fig. 6. The variation of (a) arm lengths, (b) difference of arm lengths, (c) angles between arms and (d) velocities in the measure direction in 20 years. In this figure the labeling of spacecraft and arms is different from that of Fig. 1: we have the correspondence S/C 1 → S/C 3 (Fig. 1), S/C 2 → S/C 2 (Fig. 1), S/C 3 → S/C 1 (Fig. 1), and Arm 1 → Arm 1 (Fig.1), Arm 2 → Arm 3 (Fig.1), Arm 3 → Arm 2 (Fig.1). Initial time is at noon, June 21, 2028 (JD2461944.0).

In the next step, we trim the S/C eccentricities to be nearly circular, until initial heliocentric distance getting close to the perihelion or the aphelion distance.

In the general case, after application of these two steps once, the ASTROD-GW requirements would still not be satisfied. Repeated applications of period and eccentricity optimizations are needed to satisfy the requirements. The initial conditions obtained after period optimization, and after all optimizations are listed for comparison in column 4 and 5 of Table 1 of Ref.'s 68 in the J2000 equatorial solar-system-barycentric coordinate system. The variation of arm lengths, difference of arm lengths, angles between arms, and velocities in the line-of-sight direction are drawn in Figure 6 for 20 years.[68] The index label of arms is defined by the index of the opposite spacecraft, and the index label of angle is the same as the index of spacecraft.

### 6.2. Inclination $\lambda = 1°$ case

Now consider orbit configuration with inclination $\lambda =1°$. We start with the same initial conditions at noon, June 21, 2025 (JD2460848.0) as those in Ref. 64 and 65 with equilateral distribution of S/C in the ecliptic plane, and rotate orbits by $\lambda$ using Eq. (8) to obtain the initial choice of initial conditions for this case.[57] When we calculate the arm lengths, the arm S/C 2-S/C 3 remain almost constant while for the other two arms, the 10-year trend of S/C 1-S/C 2 increases linearly with yearly ripples and the 10-year trend of S/C 1-S/C 3 decreases linearly almost symmetrically. Therefore we adjust the two components of the initial velocity of S/C 1 in the ecliptic plane and leave other things fixed to obtain an optimization. After optimization, the arm lengths and their differences are shown in Fig. 7.

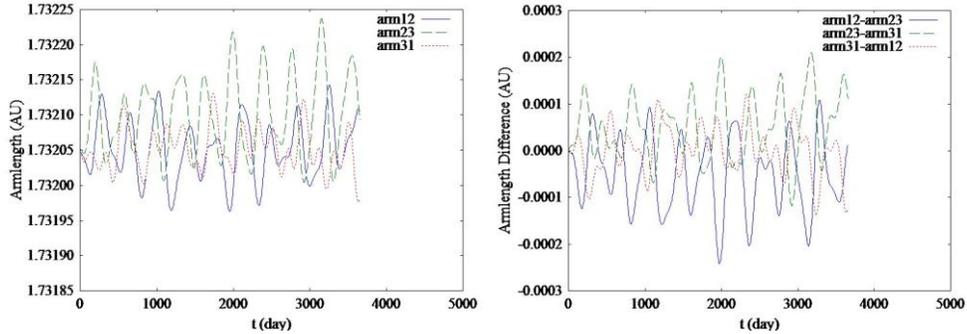

Fig. 7. Arm lengths (right), and arm lengths differences (left) for the case of 1° inclination after adjustment of initial velocity of S/C 1. Initial time is at noon, June 21, 2025 (JD2460848.0).

The arm length variations are within ± 0.00015 AU for all 3 arms. The arm length differences are within ± 0.00025 AU for all 3 interferometers.

We are currently working on optimization of orbit configurations for a period of 20 years and inclinations between 1° to 3°.

## 7. Deployment of ASTROD-GW Formation

To deploy the formation, we have given a preliminary design of the transfer orbits of the spacecraft from the separations of the launch vehicles to the mission orbits.[57]

Each spacecraft is propelled by a high efficient propulsion module for large delta-V maneuvers and for delivery to the destination. This module is to be separated when the destination is achieved.

In the mission study of ASTROD I, the ASTROD I spacecraft is given an appropriate delta-V before the last stage of launcher separation in the LEO (Low Earth Orbit) and is injected directly to the solar orbit going geodetic to Venus swing-by. We can use the same strategy to launch the ASTROD-GW spacecraft directly into the solar transfer orbits near the designated Hohmann transfer orbits or Venus swing-by orbit. This way, the only major delta V needed for each spacecraft to reach the destination occurs near the destination to boost the spacecraft to stay near the destined Lagrange point. In Table 1, we list types of transfer orbits, transfer times, the values of solar transfer delta-V and propellant mass ratio for 3 ASTROD-GW spacecraft. The propellant mass ratios are around 0.5-0.55, 0.280 and 0.47 for spacecraft 1, 2 and 3. The total masses correspond to a dry mass of 500 kg are 1111-1266 kg, 723 kg, and 1035 kg for 3 S/C respectively (including the propellant and the propulsion module with mass of 10% of the propellant).

Table 1. Estimated Delta-V and Propellant Mass Ratio for Solar transfer of S/C

| S/C (Destination) | Transfer Orbit | Transfer Time | Solar Transfer Delta-V after injection from LEO to solar transfer orbit | Solar Transfer Propellant Mass Ratio (Isp=320 s) |
|---|---|---|---|---|
| 1 (near L3) | Venus flyby transfer | 1.3-1.5 yr | 2.2-2.5 km/s | 0.50-0.55 |
| 2 (near L4) | Inner Hohmann, 2 Revolutions | 1.833 yr | 1.028 km/s | 0.280 |
| 3 (near L5) | Outer Hohmann, 1 Revolutions | 1.167 yr | 2 km/s | 0.47 |

Further studies on the optimizations of deployment from separation of launcher(s) for the orbit configurations with inclinations between 1° to 3° and for a period of 20 years are ongoing.

## 8. Time Delay Interferometry (TDI)

For the laser-interferometric antennae for space detection of GWs like ASTROD, ASTROD-GW, LISA and NGO/eLISA, the arm lengths vary according to orbit dynamics. In order to attain the requisite sensitivity, laser frequency noise must be suppressed below the secondary noises such as the optical path noise, acceleration noise etc. For suppressing laser frequency noise, time delay interferometry to match the optical path length of different beams is needed. The better optical path lengths match, the better is to cancel the laser frequency noise and the easier to achieve the requisite sensitivity. In case of exact match, the laser frequency noise is fully cancelled, as in the original Michelson interferometer.

In the study of ASTROD mission concept, time-delay interferometry was first used.[20,69,70] In the deep-space interferometry, long distance is invariably involved and laser light is attenuated to great extent at the receiving spacecraft. To transfer the laser light back or to another spacecraft, amplification is needed. The procedure is to phase lock the local laser to the incoming weak laser light and to transmit the local laser back to another spacecraft. We have demonstrated in the laboratory the phase locking of a local oscillator with 2 pW laser light.[71,72] Dick *et al.*[73] have demonstrated phase locking to 40 fW incoming weak laser light. The power requirement feasibility for ASTROD-GW is met with these developments. In the 1990s, we used the following two time-delay interferometry configurations during the study of ASTROD interferometry and obtain numerically the path length differences using Newtonian dynamics:[20,69,70]

(i) Unequal arm Michelson TDI Configuration:
Path 1: S/C 1 → S/C 2 →S/C 1 → S/C 3 → S/C 1,
Path 2: S/C 1 → S/C 3 →S/C 1 → S/C 2 → S/C 1.

(ii) Sagnac TDI configuration:
Path 1: S/C 1 → S/C 2 → S/C 3 → S/C 1,
Path 2: S/C 1 → S/C 3 → S/C 2 → S/C 1.

Time delay interferometry has been worked out for LISA much more thoroughly since 1999.[74,75] First-generation and second-generation TDIs are proposed. In the first generation TDIs, static situations are considered, while in the second generation TDIs, motions are compensated to certain degrees. The two configurations considered above are first generation TDI configurations in the sense of Armstrong, Estabrook and Tinto.[74,75] Fig. 8 shows the path length differences of ASTROD-GW for these two TDI configurations (with the S/C labeling switched between 1 and 3 in Fig. 1) calculated numerically using CGC 2.7 ephemeris. We shall not review more about these historical developments here, but refer the readers to the excellent review of Tinto and Dhurandhar.[75]

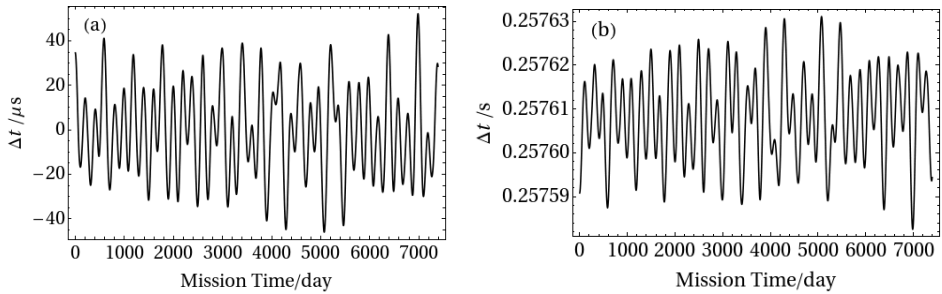

Fig. 8. The path length difference of two optical paths of ASTROD-GW for (a) unequal-arm Michelson TDI configuration (X observable) [*a*, *b*] and (b) Sagnac configuration.

We have simulated the time delay numerically using CGC2.7 ephemeris framework for ASTROD-GW,[66-68,76] LISA[77] and NGO/eLISA.[78] To conform to the ASTROD-GW planning, we have worked out as an example a set of 20-year optimized mission orbits of ASTROD-GW spacecraft starting at June 21, 2028 (Sec. 6.1.2 and Fig. 6), and calculate the residual path differences in the first and second generation time delay interferometry. For the numerical evaluation, we take a common receiving time epoch for both beams; the results would be very close to each other numerically if we take the same starting time epoch and calculate the path difference. The results of this calculation for the unequal arm Michelson TDI configuration and for the Sagnac TDI configuration are shown in Fig. 8. If we refer to the path S/C1 → S/C2 →S/C1 as *a* and the path S/C1 →

S/C3 → S/C1 as *b*, then the difference of Path 1 minus Path 2 for the unequal-arm Michelson can be denoted as *ab-ba* ≡ [*a*, *b*].[79] This configuration is called the X observable for LISA.[74,75]

LISA, NGO/eLISA and ASTROD-GW might need second generation TDI for meeting the noise requirement of laser noise suppression. Besides the Unequal-Arm Michelson and Sagnac TDI configurations/observables, our calculations of the first order TDI variables for ASTROD-GW include Symmetrized Sagnac, Relay, Beacon, and Monitor configurations/observables.[66,67] These configurations/observables constitute all the six-link and eight-link observables for triangular formation.[80] We also calculate many second order TDI configurations/observables.[66,67,76] For all first and second generation TDI calculated for ASTROD-GW, their interferometric optical path differences are less than 150 ns (50 m) and are well within the requirement for the ASTROD-GW mission.[67,68,76]

## 8.1. Second-generation TDI for two arm case of ASTROD-GW

In the following, we discuss the path length differences of second-generation TDIs calculated numerically for the two-arm case of ASTROD-GW.

The second-generation TDIs obtained by Dhurandhar, Nayak and Vinet[79] are listed in degree-lexicographic (binary) order as follows:

(i) n=1, [*ab*, *ba*] = *abba* – *baab* ;
(ii) n=2, [$a^2b^2$, $b^2a^2$], [*abab*, *baba*], [$ab^2a$, $ba^2b$];
(iii) n=3, [$a^3b^3$, $b^3a^3$], [$a^2bab^2$, $b^2aba^2$], [$a^2b^2ab$, $b^2a^2ba$], [$a^2b^3a$, $b^2a^3b$],
    [$aba^2b^2$, $bab^2a^2$], [*ababab*, *bababa*], [$abab^2a$, $baba^2b$], [$ab^2a^2b$, $ba^2b^2a$],
    [$ab^2aba$, $ba^2bab$], [$ab^3a^2$, $ba^3b^2$].

For the numerical evaluation, we take a common receiving time epoch for both beams and calculate their path difference. Fig. 9 shows the numerical results for the second-generation n = 1 configuration and three second-generation n = 2 TDI configurations. It is interesting to note that for the last case in Fig. 9, namely [$ab^2a$, $ba^2b$], the path differences are in the range -140 to 120 ps, much below the others. This is the result of cancellation of higher order terms in the time derivatives of the arm lengths. Specifically, there is greater symmetry in this combination in which the second derivative terms also cancel out.

In Table 2, we compile and compare the resulting differences for the 14 TDIs listed in the beginning of this subsection due to different nominal arm lengths for various mission proposals -- NGO/eLISA, an NGO-LISA-type mission with a nominal arm length of 2 × $10^6$ km, LISA and ASTROD-GW. For NGO/eLISA, an NGO-LISA-type mission and LISA, the starting epoch is at noon, January 1st, 2021 and we have used the CGC 2.7 ephemeris for doing the calculation for the 1000-day optimized mission orbit configurations obtained in Ref. 77 and Ref. 78. All the mission proposals satisfy their respective requirements well for these second generation TDI observables.

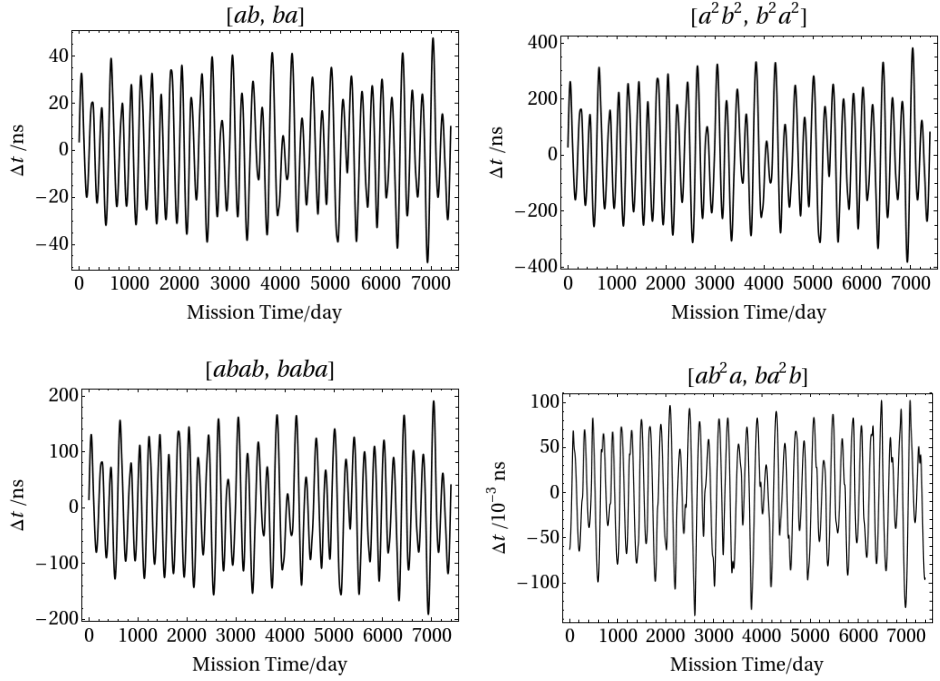

Fig. 9. The path length differences (as a function of epoch) of two optical paths of n = 1 TDI configuration ([*ab*, *ba*], upper left diagram) and three n = 2 TDI configurations.

Table 2. Comparison the resulting differences due to arm lengths for various mission proposals -- NGO/eLISA, an NGO-LISA-type mission with a nominal arm length of $2 \times 10^6$ km, LISA and ASTROD-GW.

| TDI configuration | | TDI path difference $\Delta L$ | | | |
|---|---|---|---|---|---|
| | | NGO/eLISA [ps] | NGO-LISA-type With $2 \times 10^6$ km arm length [ps] | LISA [ps] | ASTROD-GW [ns] |
| n=1 | [*ab*, *ba*] | -1.5 to +1.5 | -11 to +12 | -70 to +80 | -50 to +50 |
| n=2 | [$a^2b^2$, $b^2a^2$] | -11 to +12 | -90 to +100 | -600 to +650 | -400 to +400 |
| | [*abab*, *baba*] | -6 to +6 | -45 to +50 | -300 to +340 | -200 to +200 |
| | [$ab^2a$, $ba^2b$] | -0.0032 to +0.0034 | -0.0036 to +0.004 | -0.015 to +0.013 | -0.14 to +0.11 |
| n=3 | [$a^3b^3$, $b^3a^3$] | -40 to +42 | -300 to +320 | -2,000 to +2,200 | -1,300 to +1,300 |
| | [$a^2bab^2$, $b^2aba^2$] | -30 to +32 | -220 to +260 | -1,500 to +1,800 | -1,100 to +1,100 |
| | [$a^2b^2ab$, $b^2a^2ba$] | -22 to +24 | -160 to +180 | -1,000 to +1,300 | -750 to +750 |
| | [$a^2b^3a$, $b^2a^3b$] | -13 to +14 | -100 to +110 | -600 to +750 | -450 to +450 |
| | [$aba^2b^2$, $bab^2a^2$] | -22 to +24 | -160 to +180 | -1,000 to +1,300 | -750 to +750 |
| | [*ababab*, *bababa*] | -13 to +14 | -100 to +110 | -600 to +750 | -450 to +450 |
| | [$abab^2a$, $baba^2b$] | -4.5 to +4.8 | -32 to +38 | -200 to +250 | -150 to +150 |
| | [$ab^2a^2b$, $ba^2b^2a$] | -4.5 to +4.8 | -32 to +38 | -200 to +250 | -150 to +150 |
| | [$ab^2aba$, $ba^2bab$] | -4.8 to +4.5 | -38 to +32 | -250 to +200s | -150 to +150 |
| | [$ab^3a^2$, $ba^3b^2$] | -15 to +13 | -110 to +100 | -750 to +600 | -450 to +450 |
| Nominal arm length | | 1 Gm (1 Mkm) | 2 Gm | 5 Gm | 260 Gm |
| Requirement on $\Delta L$ | | 10 m (30,000 ps) | 20 m (60,000 ps) | 50 m (150,000 ps) | 500 m (1,500 ns) |

## 9. Payload Concept

The ASTROD-GW mission consists of three spacecraft, each of which orbits near one of the Sun-Earth Lagrange points (L3, L4 and L5), such that the array forms an almost equilateral triangle. The 3 spacecraft range interferometrically with one another with an arm length of about 260 million kilometers. Each spacecraft carries a payload of two proof masses, two telescopes, two 1-2 W lasers, a weak light handling system, a clock or a molecule line stabilization system for lasers, an absolute laser metrology system, and a drag-free system.[1,14-16] The LISA Pathfinder,[78] ASTROD I,[7] DECIGO Pathfinder[55] and NGO/eLISA[18] will be flying before ASTROD-GW. Much of the payload will be implemented already before the ASTROD-GW payload implementation. We will mainly address the issues more specific to ASTROD-GW.

*Weak light phase locking and handling*: For a distance of 260 Gm (1.73 AU), we need to phase lock a local laser to 100 fW incoming light to amplify and manipulate it. For 100 fW ($\lambda$ = 1064 nm) weak light, there are $5 \times 10^5$ photons/s. This would be good for 100 kHz frequency tuning. In Tsing Hua University, 2 pW weak-light phase-locking with 0.2 mW local oscillator has been demonstrated.[71,72] In JPL (Jet Propulsion Laboratory), Dick *et al*.[73] have achieved offset phase locking to 40 fW incoming laser light. The future development should be focused on frequency-tracking, modulation-demodulation and coding-decoding to make it a mature experimental technique. This is also important for the deep space optical communication.

*Drag-free system design and development*: With the precision requirement of this mission, drag-free is a must. Drag-free system consists of a high precision accelerometer/inertial sensor to detect non-drag-free motions and a micro-thruster system to do the feedback to keep the spacecraft drag-free. The drag-free requirement for the ASTROD-GW in the frequency range above 0.1 mHz is similar to LISA and NGO/eLISA. This sensitivity is a goal to be largely demonstrated by LISA Pathfinder in 2015.[75] In the frequency range 100 nHz to 0.1 mHz, the present goal of ASTROD-GW is to follow enhanced LISA sensitivity goal down to 10 μHz and extrapolate it further to 100 nHz. This requires position sensing noise to be flat down to 100 nHz and gravity acceleration due to spacecraft to be small and modeled to the required level at low frequencies. The self-gravity-acceleration needs to be stable or subtracted in real time. An absolute laser metrology system to monitor positions of major mass distribution in the S/C will be implemented to do this. After a careful analysis of noises of test mass and capacitive sensing, Bender[28] suggested a specific sensitivity goal at frequencies down to 3 μHz which contained a milder "reddening factor" (compared to later adopted MLDC "reddening factor"[27]). With monitoring the gap of capacitive sensing and the positions of major mass distribution in the spacecraft, we hope that this factor may be alleviated to certain extent. To completely drop the factor or to go beyond, one may need to go to optical sensing and optical feedback control.[29-37] As to the accelerometer/inertial sensor design of ASTROD, an absolute laser metrology system is required to push the noise down, in particular in the lower frequency region. Since ASTROD-GW would be launched after NGO/eLISA, ASTROD-GW can learn from the drag-free technologies developed for LISA. In addition, ASTROD needs monitoring the positions of various parts of the spacecraft, to facilitate gravitational modeling.[31,37]

*Micro-thruster system*: For drag-free feedback control, micro-thrusters are needed. Field Emission Electric Propulsion (FEEP) system with its high specific thrust is a good candidate for the micro-thruster system. The sensitivity of FEEP system is good and is in the μN range. The main issue for FEEPs is lifetime. Due to technical problems during the development of the FEEP technology, the cold gas thrusters have become the alternative choice. The GAIA mission will carry cold gas thrusters for the AOCS (Attitude and orbit control system).[82] MICROSCOPE and LISA Pathfinder will be equipped with cold gas thrusters based on the GAIA thrusters. The main disadvantage of cold gas thrusters compared to FEEPs is the higher mass per delta-V. The total mission duration is limited by the amount of propellant stored in the tanks. Therefore the FEEP technology would be preferred if it is available at a later time for ASTROD-GW.

*Laser system*: Nd:YAG Non-planar ring oscillators pumped by laser diodes are available with output power of 2 W for use. The frequency noises must suppressed to very low level. The strategy is like the one adopted by NGO/eLISA using pre-stabilization, arm locking[18] and TDI (Sec. 8).

*Laser frequency standard/Clock*: It is desirable to have light-weight precision space clock and/or absolute-stabilized laser with accuracy to $10^{-15}$ or better for ASTROD-GW mission. In view of the present developments of ion frequency standards and optical standards, this looks feasible in the time frame considered for a 2026-2027 launch.

Space optical clocks and optical comb frequency synthesizer technologies are important in the realization and simplification of the ASTROD-GW target sensitivity at low frequency. Another use of the optical clock and optical comb frequency synthesizer is to calibrate the optical metrology for ASTROD-GW. This is important for the laser metrology inertial sensor and for monitoring distances inside spacecraft, to correct local gravity changes due to, for example, thermal effects. All these measurements use lasers as standard rods. They need to be calibrated using optical frequency standards or absolutely stabilized laser frequency standard referenced to an atomic or molecular line. The advent of optical clocks and optical combs in space may possibly simplify the experimental design of ASTROD-GW mission.

At present, optical clocks in the laboratory have reached a fractional inaccuracy of $8.6 \times 10^{-18}$ (an uncertainty equivalent to one second in 4 billion years);[83] and they are improving. Clocks of this accuracy level or better can be used for exquisitely sensitive measurements of gravity, motion, and inertial navigation. The use of this kind of clocks certainly will facilitate the detection of the lower frequency GWs and stimulate the needs of re-design the implementation schemes of the lower frequency GW detection.

*Absolute laser metrology system*: With an ultraprecise laser frequency standard/clock, an absolute laser metrology system can be built to monitor the positions of various parts of the spacecraft to facilitate gravitational modeling.

*Radiation monitor*: A small radiation detector onboard the spacecraft will monitor test-mass charging of the inertial drag-free sensors. This radiation monitor can also be used for measuring solar energetic particles (SEPs) and galactic cosmic rays (GCRs) in the area of solar and galactic physics with corresponding applications to space weather.[84,85]

## 10. Summary and Outlook

We have reviewed the scientific goals, the frequency sensitivity spectrum, the basic orbit configuration, the optimization of the configuration/constellation, the angular resolution, deployment of the formation, the TDI and the payload for ASTROD-GW. *The basic orbit configuration of ASTROD-GW is unique in the Sun-Earth system for the detection of GWs in the low frequency band.* This is because Sun-Earth Lagrange points L4, L5 are stable, and L3 is quasi-stable with a time scale over 50 years. The three spacecraft in circular orbits around the Sun near these three Lagrange points are (quasi-)stable for 20 years and can form a nearly equilateral triangle geodetically. Formation inclined to the ecliptic plane with angular precession of half sidereal year period is desired for resolving the angular ambiguity. This will also resolve the angular resolution in polar region of the sky. For the inclined formation of 1° inclination, we optimized the orbit configuration to have arm length variations less than $\pm 10^{-4}$ (arm length differences less than $\pm 1.5 \times 10^{-4}$) fractionally in ten years. Relative Doppler velocities between spacecraft are within $\pm 3$ m/s for easier frequency heterodyning.

Although how well one can improve on the lower frequency part of LISA and NGO/eLISA accelerometer/inertial sensor noise over the MLDC formula is not clear, *the strain sensitivity of ASTROD-GW in the lower frequency part is better than LISA and NGO/eLISA by 52 and 260 times respectively due to longer arm length if one assumes the same accelerometer/inertial sensor noise. ASTROD-GW has the best sensitivity in the 0.1 nHz to 1 mHz frequency band bridging the gap between LISA, NGO/eLISA and PTAs.* From this vintage point, ASTROD-GW is unique in complementing NGO/eLISA and PTAs in the exploration of black hole co-evolution with galaxies, the determination of the equation of state of dark energy, testing relativistic gravity and probing the inflationary physics. Efforts in minimizing the accelerometer/inertial sensor noise over the MLDC formula will strengthen these goals. Deployment of ASTROD-GW formation is studied with deployment time less than 1.8 years and propellant mass ratio less than 0.55. This is within the practical range of launcher implementation.

Now we list important issues for further studies in order to realize and sharpen our expectations:

(i) Manipulating weak light,
(ii) Improvement of low-frequency acceleration noise,
(iii) Fourier spectrum of perturbations due to celestial bodies in the solar system and precision needed to know the positions of solar-system bodies in order to separate this spectrum from GW spectrum,
(iv) Further studies in optimizing deployment delta-V and propellant ratio (common problem to DECIGO/BBO and some other future missions to come),
(v) Optimizing the inclination angle of the ASTROD-GW constellation,
(vi) Extraction of GW signals based on precise numerical orbits,
(vii) Further studies in the angular resolution of GW sources of ASTROD-GW,
(viii) Separation of weak lensing effects from GW signals.

Space missions using optical devices will be important in testing relativistic gravity and measuring solar-system parameters. Laser Astrodynamics in the solar system envisages ultra-precision tests of relativistic gravity, provision of a new eye to see into the solar interior, precise measurement of $\dot{G}$, monitoring the solar-system mass loss, and detection of low-frequency gravitational waves to probe the early Universe and study strong-field black hole physics together with astrophysics of binaries. One spacecraft and multi-spacecraft mission concepts --- ASTROD I, ASTROD-GW and Super-ASTROD are in line for more thorough mission studies. Mission studies along this line will be fruitful.

GW experiments become a world focus and we are very close to a detection.[25] Interests in GW detection have also sprung up in Asia. With (i) 3 km arm-length underground cryogenic laser-interferometric GW detector KAGRA/LCGT[86] under construction and the 1000 km Fabry-Perot space interferometer DECIGO[54,55] under active planning in Japan, (ii) the startup of IndIGO and interests in space GW detection in India,[25,87,88] and (iii) the five-hundred-meter aperture spherical radio telescope (FAST) under construction[89] and the activities to initiate/join collaboration in space GW projects in China[90] together with (iv) PTAs actively searching for GWs,[43,44] (v) adLIGO,[91] adVIRGO[92] and SKA[93] under construction, (vi) LPF scheduled for launch in 2015,[81] and (vii) ET[94,95] and NGO/eLISA[18] under active planning, we are looking forward to an era with important advances in GW astronomy, cosmology and physics.[25]

## Acknowledgements

I am grateful to Professor Bala Iyer, Director Ravi Subrahmanyan and staff members in Raman Research Institute (RRI) for their great hospitality and support during my visit in 2011 and for organizing the ASTROD 5 Symposium in 2012. I am also grateful to all members of the ASTROD/ASTROD I/ASTROD-GW study projects -- in particular, Gang Wang and An-Ming Wu -- for their works and encouragements which make this presentation possible. I would like to thank C. K. Mishra, M.-L. Tong and K. Yagi for pointing out a discrepancy between the sensitivity formula and the figure which I have repeatedly overlooked. I would also like to thank Bala Iyer and Dah-Wei Chiou for a critical reading of the manuscript. For writing this review, I thank the National Science Council (Grant No. NSC101-2112-M-007-007) and the National Center of Theoretical Sciences (NCTS) for support.